\documentclass[aps,pre,twocolumn,amssymb,amsmath,showpacs,showkeys,superscriptaddress,floatfix,a4paper]{revtex4}

\usepackage[final]{graphicx}

\usepackage{dcolumn}

\begin{document}

\title{Properties of Quantum Systems via Diagonalization of Transition\\
Amplitudes II: Systematic Improvements of Short-time Propagation}
\author{Ivana Vidanovi\'c}
\author{Aleksandar Bogojevi\'c}\email[E-mail: ]{aleksandar.bogojevic@scl.rs}
\author{Antun Bala\v z}
\author{Aleksandar Beli\'c}
\affiliation{Scientific Computing Laboratory, Institute of Physics Belgrade, Pregrevica 118, 11080 Belgrade, Serbia}
\homepage[Home page: ]{http://www.scl.rs/}

\begin{abstract}
In this paper, building on a previous analysis \cite{pqseeo1} of exact diagonalization of the space-discretized evolution operator for the study of properties of non-relativistic quantum systems, we present a substantial improvement to this method. We apply recently introduced effective action approach for obtaining short-time expansion of the propagator up to very high orders to calculate matrix elements of space-discretized evolution operator. This improves by many orders of magnitude previously used approximations for discretized matrix elements and allows us to numerically obtain large numbers of accurate energy eigenvalues and eigenstates using numerical diagonalization. We illustrate this approach on several one and two-dimensional models. The quality of numerically calculated higher order eigenstates is assessed by comparison with semiclassical cumulative density of states.
\end{abstract}

\preprint{SCL preprint}
\maketitle

\section{Introduction}
\label{sec:intro}

In first paper in this series \cite{pqseeo1} we analyzed in detail the earlier introduced method \cite{sethia, sethiacpl1, sethiajcp, sethiacpl2} for studying properties of quantum systems based on the diagonalization of real-space discretized evolution operator, as well as the errors associated with the discretization process. This analysis provided us with a better understanding of this method that can be used to numerically calculate energy eigenvalues and eigenstates of few-body physical systems. We have shown that errors due to the finite discretization step $\Delta$ used for space discretization vanish exponentially with $1/\Delta^2$ for short time of propagation. This highly outperforms the usual (polynomial in $\Delta^2$) behavior of errors in approaches with diagonalization of space-discretized Hamiltonians \cite{spacedisrev,d,spacedis1,spacedis2}. In addition, derived analytic estimates for discretization errors provide an easy way for taking into account such errors and eliminating them in practical applications. For these reasons, the approach \cite{sethia} now becomes the preferred method to study systems with few degrees of freedom, for which it can be efficiently and straightforwardly implemented. In our previous paper we have also analyzed estimates for errors due to the introduction of the space cutoff $L$, providing simple criterion for assessing and eliminating errors of this type.

In the previous paper we have also demonstrated that time of propagation $t$, a parameter introduced by this method, is a source of a new type of error that comes about from using short time approximations. This problem was not addressed at all in Ref.~\cite{sethia}. It has recently been discussed \cite{k1, k2, k3} and will be the main focus of the present paper. Errors associated with the time of evolution parameter $t$ must be carefully taken into account and may substantially limit the precision of numerical calculation in the diagonalization method. In this paper we address this problem by applying the recently introduced effective action approach \cite{b,prl-speedup,prb-speedup,pla-euler,pla-manybody} to systematically improve approximations for transitions amplitudes. This in turn leads to the many orders of magnitude decrease of errors in obtained energy eigenvalues, as shown in this paper. We demonstrate on several lower-dimensional models how use of higher-order effective actions significantly reduces numerical errors and systematically improves obtained energy eigenvalues and eigenstates.

Together with the results from our previous paper, this paper completes the analysis of the method based on the diagonalization of transition amplitudes, providing us with necessary analytical knowledge to estimate errors of all types associated with this method and to numerically very accurately calculate large numbers of energy eigenvalues and eigenstates. This invites various applications of the method to the study of few-body quantum systems, some of which are discussed throughout the paper.

The text is organized as follows: in Section~\ref{sec:effective} we briefly review the effective action approach and demonstrate how it can be used for numerical calculation of transition amplitudes. In Sections~\ref{sec:d1}  and \ref{sec:d2} we apply the exact diagonalization method \cite{pqseeo1,sethia} improved by the use of effective actions to the numerical study of several one and and two-dimensional models. In these sections we also show how the number of reliable energy eigenvalues can be estimated using comparison of numerically obtained results with semiclassical cumulative density of states for higher-lying eigenstates. Section~\ref{sec:conclusions} gives our concluding remarks and some relevant applications of this approach.

\section{Effective actions}
\label{sec:effective}

To introduce the notation, we first briefly outline the diagonalization method \cite{sethia}, presented in more detail in Section 2 of our previous paper \cite{pqseeo1}. After discretizing the continuous space and replacing it with a grid defined by a discretization step $\Delta$, all the quantities are defined only on a discrete set of coordinates $x_n=n\Delta$, where $n\in\mathbb{Z}$ is any integer number. For a physical system with Hamiltonian $\hat{H}$, the evolution operator (in the imaginary time formalism) is defined as $\exp(-t\hat{H})$, where $t$ is the time of evolution. Transition amplitudes are defined as
\begin{equation}
A(x,y;t)=\langle x|e^{-t\hat{H}}|y\rangle\, ,
\end{equation}
and give the discretized evolution operator matrix elements
\begin{equation}
A_{nm}(t)=\Delta^d\cdot A(n \Delta, m\Delta;t)\, ,
\end{equation}
where $d$ is the number of spatial dimensions. The eigenvectors of such a matrix correspond to the space-discretized eigenfunctions of the original Hamiltonian, while the eigenvalues are related to the eigenvalues of the Hamiltonian and can be written as
\begin{equation}
e^{-tE_k(\Delta,L, t)}\, ,
\label{eq:en}
\end{equation}
where we emphasize the dependence of the numerically calculated eigenvalues on all discretization parameters. The number of obtained eigenvalues and eigenstates is equal to the
linear size of matrix $A$, which has to be finite when we represent any physical system on the computer. Typically, we restrict the range of indices $n$, $m$ to the finite interval $-N\leq n, m < N$, so that the number of points in the grid is $S=(2N)^d$. Note that the range can be adjusted so that the size $S$ is an odd number. In Eq.~(\ref{eq:en}) we have also introduced the space cutoff $L$, which corresponds to the restriction on the range of grid-point indices $n,m$, and is given by $L=N\Delta$.

As we can see, the precise calculation of transition amplitudes is essential for practical applications of this method. In Ref.~\cite{sethia} all calculations are based on the naive approximation for transition amplitudes
\begin{equation}
A^{(1)}(x,y;t)\approx\frac{1}{(2\pi t)^{d/2}}\, e^{-\frac{(x-y)^2}{2 t}-t \frac{V(x)+V(y)}{2}}\, ,
\end{equation}
which yields energy eigenvalues correct only to order $O(t)$, and is for this reason designated by $A^{(1)}$.  If one uses the naive approximation for transition amplitudes, then times of propagation must be very short for errors to be small enough. Practically, even for short times of propagation, such errors are always much larger than the errors due to discretization, and therefore significantly limit the applicability of the method. In addition to this, the results obtained in our previous paper \cite{pqseeo1} on exactly solvable models suggest that longer times of propagation generally give smaller errors in the diagonalization approach. The trade-off between these effects and its implications on numerical results have been documented in \cite{sethia}.

\begin{figure}[!t]
\centering
\includegraphics[width=8.5cm]{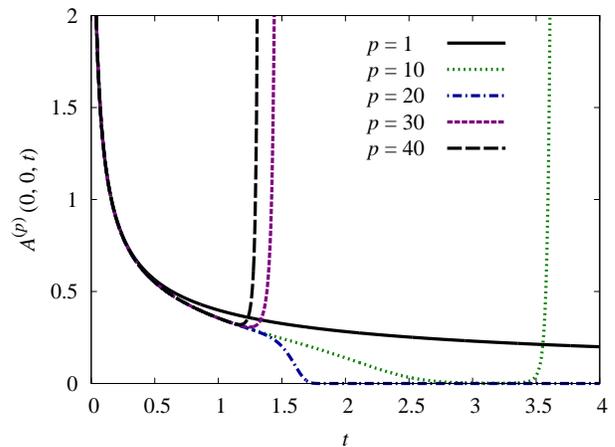}
\caption{(Color online) Transition amplitude $A^{(p)}(0,0;t)$ as a function of the time of propagation $t$, calculated analytically using different levels $p$ of the effective action. The plot is for the quartic anharmonic potential $V(x)=\frac{1}{2}M\omega^2x^2+\frac{g}{24} x^4$, with parameters $M=\omega=1$, $g=10$.}
\label{fig:amplitude2}
\end{figure}

To address this, in principle one can use Monte Carlo simulations \cite{ceperley,boninsegni} to calculate amplitudes $A$ to high precision. Although this can effectively resolve the problem in many cases, it is often numerically very expensive. More importantly, resorting to the use of Monte Carlo practically limits further analytical approaches. We will instead use the recently introduced effective action approach \cite{b,prl-speedup,prb-speedup,pla-euler,pla-manybody} that gives closed-form analytic expressions  $A^{(p)}(x,y;t)$ for transition amplitudes which converge much faster,
\begin{equation}
A^{(p)}(x,y;t)=A(x,y;t)+O(t^{p+1}/t^{d/2})\, ,
\label{eq:impcon}
\end{equation}
where $p$ is an integer number corresponding to the order of the effective action used, i.e. order of energy eigenvalue errors $t^p$. For a general many-body theory effective actions up to $p=10$ have been derived, while for a specific models much higher values can be obtained, e.g. for the anharmonic oscillator and other polynomial interactions, for which effective actions have been calculated up to $p=144$. So, if $p$ is high enough, it is sufficient that the time of evolution is less than the radius of convergence of the above series ($t<\tau_c\sim 1$) and errors in calculated values of transition amplitudes will be negligible. This is illustrated in Fig.~\ref{fig:amplitude2} for the case of a quartic anharmonic oscillator. The use of high-order expansion in the time of propagation of amplitudes will allow us to use times of evolution up to $\tau_c$, which are much longer than the typical times one can use with the naive ($p=1$) amplitudes. At the same time, the expansion up to very high orders substantially decreases the errors associated with $t$, and may practically eliminate them.

The analytic expressions for higher-order approximations for transition amplitudes are based on the notion of effective actions, which are introduced by casting the solution of the time dependent Schr\"odinger equation for the transition amplitude in the form
\begin{equation}
A(x,y;t)=\frac{1}{(2\pi t)^{d/2}}\, e^{-\frac{(x-y)^2}{2 t}-tW\left(\frac{x+y}{2},x-y;t\right)}\, ,
\label{eq:form}
\end{equation}
where $W(x,\delta;t)$ is the effective potential, with the following boundary behavior:
\begin{equation}
\lim_{t\rightarrow 0}W(x,\delta;t)=V(x)\, .
\end{equation}
As shown previously, the effective potential $W(x,\delta;t)$ is regular in the vicinity of $t=0$, enabling us to represent it in the form of a power series in short time of propagation $t$. The coefficients in this series are functions of the potential and its derivatives. The truncation of the series for the effective potential up to order $t^{p-1}$, designated by $W^{(p-1)}(x,\delta;t)$, gives the expansion of the transition amplitude accurate to $t^{p+1}/t^{d/2}$,
\begin{equation}
A^{(p)}(x,y;t)=\frac{1}{(2\pi t)^{d/2}} e^{-\frac{(x-y)^2}{2 t}-tW^{(p-1)}\left(\frac{x+y}{2},x-y;t\right)} .
\end{equation}
The analytic expressions for higher-order effective actions therefore yield analytic approximations for amplitudes with the convergence behavior given by Eq.~(\ref{eq:impcon}). We emphasize that although the structure of the effective action solution form (\ref{eq:form}) is motivated by the path integral formalism, the expression for amplitudes obtained in the above approach contains§ no integrals and can be used straightforwardly as long as the time of propagation is below the radius of convergence of the short-time series expansion.

For the exactly solvable case of a harmonic oscillator one finds that the radius of convergence is $\tau_c=\pi/\omega$. The radius of convergence is simply the distance in the complex time plain from the origin to the nearest singularity of the propagator. For the harmonic oscillator the singularities are located at $\pm ik\pi/\omega$, $k\in\mathbb{N}$. The consequence of these singularities is that the power series for the effective potential $W(x,\delta;t)$ converges only for $t<\tau_c$. It is often difficult to analytically determine the radius of convergence of the short time expansion of the transition amplitude. However, numerically this is a very simple problem, since outside of the radius of convergence the calculated approximative amplitudes rapidly tend to infinity (for levels $p$ for which the effective potential is not bounded from below; see Ref.~\cite{pre-ideal}) or to zero with the increase of $p$. From Fig.~\ref{fig:amplitude2} we easily estimate radius of convergence to be $\tau_c\approx 1$ for a quartic anharmonic potential  $V(x)=\frac{1}{2}M\omega^2x^2+\frac{g}{24} x^4$, with parameters $M=\omega=1$, $g=10$. Such numerical determination of the radius of convergence for a given level $p$ is always done before practical use of the effective potential. Note that we are not interested in the precise value of $\tau_c$, just in its rough estimate which will allow us to safely use times of propagation below $\tau_c$.

To conclude the section, let us stress that the effective action approach can be used only for sufficiently smooth potentials, i.e. those that have derivatives of the required order, corresponding to the level $p$ of effective action, as discussed in Ref.~\cite{b}. For potentials that do not fulfill this condition (e.g. stepwise potentials), the effective action approach cannot be directly used. However, one can replace the original potential with some of its smooth deformations, perform numerical calculations, and at the end take the limit of the deformation parameter in which the original potential is recovered. The numerical results obtained in such a way must be carefully cross-checked using other methods.

\section{Numerical results for $d=1$ models}
\label{sec:d1}

In this section we apply the approach outlined above to several $d=1$ models and demonstrate its substantial advantages for numerical studies of eigenstates of various physical systems. We numerically analyze all sources of errors present in this approach due to discretization parameters $L$ and $\Delta$, as well as the time of propagation parameter $t$. We present the obtained numerical results for energy eigenvalues and eigenstates. We also assess the quality of the obtained energy spectra through comparison with the semiclassical approximation for the density of states, which should be accurate at least for the higher regions of the spectrum.

The first model we study is the quartic anharmonic oscillator with potential
\begin{equation}
V(x)=\frac{1}{2}M\omega^2 x^2+\frac{g}{24}x^4\, .
\label{eq:quarticpot}
\end{equation}
For this potential the effective actions have been previously derived up to $p=144$ \cite{scl-speedup}, and here we will use various levels $p$ to illustrate the dependence of errors on the level $p$ used in calculations.

\begin{figure}[!b]
\centering
\includegraphics[width=8cm]{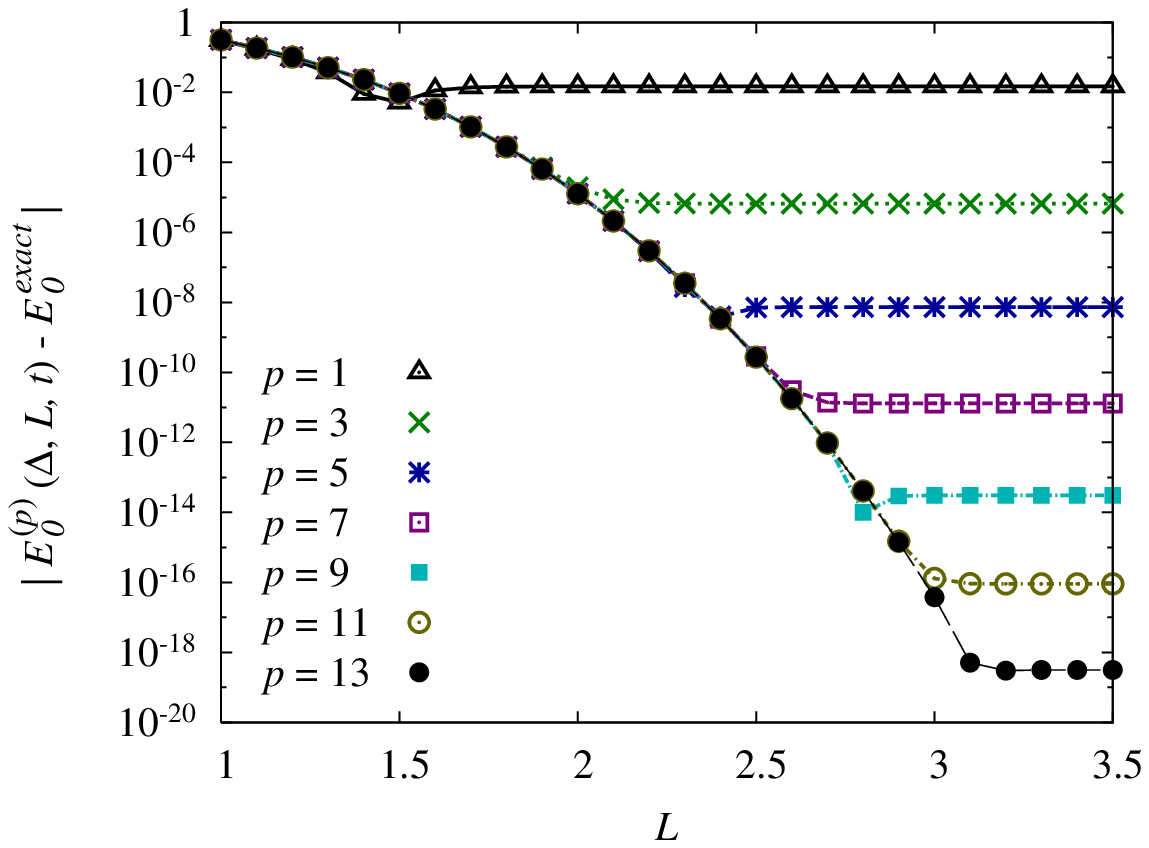}
\includegraphics[width=8cm]{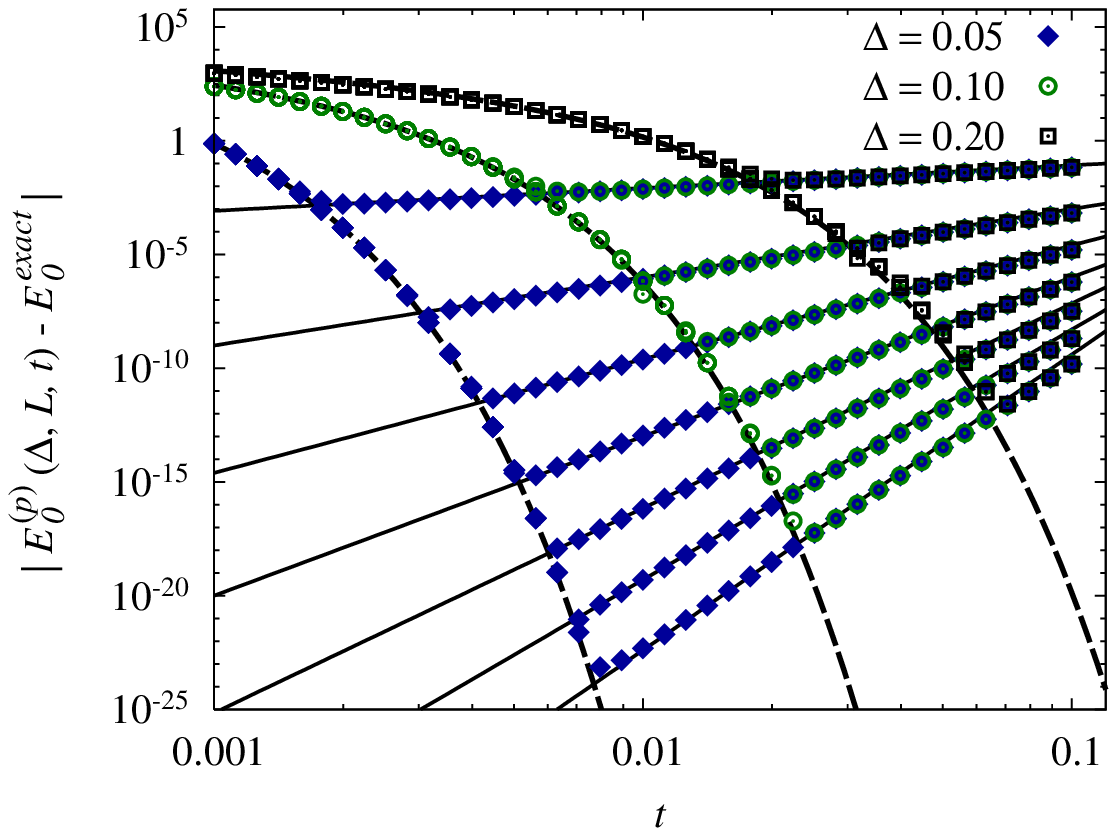}
\caption{(Color online) (a) Deviations from the ground energy $|E_0^{(p)}(\Delta,L,t) -E_0^{exact}|$ as a function of the space cutoff $L$ and (b) as a function of the time $t$. The ground energy is obtained using different levels $p=1,3,5,7,9,11,13$ (top to bottom) of the effective action for the quartic anharmonic potential, with parameters $M=\omega=1$, $g=48$, $\Delta=0.05$, $t=0.02$ on graph (a), and $L=4$ on graph (b). The exact ground energy  $E_0^{exact}=0.95156847272950001114693\ldots$ is taken from Ref.~\cite{c}. Dashed lines on the graph (b) correspond to the known discretization error \cite{pqseeo1}.}
\label{fig:quarticerr}
\end{figure}

Fig.~\ref{fig:quarticerr} presents the analysis of various errors in the ground energy calculation for a particular choice of parameters of the potential $M=\omega=1$, $g=48$. The spectrum of the potential is calculated by the numerical diagonalization of the space-discretized transition amplitude matrix. The errors are estimated using the exact value of the ground energy calculated elsewhere \cite{c} by a different technique to very high precision. The dependance of the error related to the introduction of the space cutoff $L$ is illustrated in Fig.~\ref{fig:quarticerr}a, while Fig.~\ref{fig:quarticerr}b gives the dependence of ground energy errors on the time of propagation parameter $t$ for various values of the discretization step $\Delta$. On both graphs we see the results obtained with effective actions of different levels $p$. Fig.~\ref{fig:quarticerr}b clearly shows that the errors due to the time of propagation are proportional to $t^p$, as expected when we use the effective action of the level $p$. The errors in eigenvalues are of the same order as errors in calculation of individual matrix elements, and for this reason we see the typical $t^p$ behavior of ground and higher energy eigenvalues. It is already now evident that the use of higher order effective actions increases the accuracy of numerically calculated energy eigenstates for many orders of magnitude. This is the most important contribution of this paper.

The $L$-dependence of the error is analytically known \cite{ajp1, ajp2, pqseeo1}. The saturation of errors in Fig.~\ref{fig:quarticerr}a for a given level $p$ corresponds to a maximal precision that can be achieved with that $p$, i.e. denotes the value of $L$ for which errors introduced by other sources become larger than the error due to the finite value of the space cutoff. This can be easily seen if we combine the data from both graphs. For example, the level $p=9$ effective action has the saturated value of the error of the order of $10^{-14}$. For $t=0.02$ we find that the error due to the time of propagation is of the same order if one uses sufficiently fine discretization ($\Delta=0.05$). Therefore, the saturation of errors on the left-hand graph are caused by the errors due to the time of propagation. However, if one uses discretization which is not sufficiently fine, the saturation of errors can be also caused by the discretization effects. Such effects can be also analytically estimated to be proportional to $-2\exp(-2\pi^2 t/\Delta^2)\cosh(\pi^2(k+1)t/L\Delta)/t$, as we have shown in the previous paper \cite{pqseeo1}.

\begin{table}[!t]
\begin{center}
\begin{tabular}{|c|c|c|c|}
\hline
$k$ & $E_k$&$|\Delta E_k|$ & $\delta E_k$\\
\hline\hline
$\quad 0  \quad  $ & $    0.9515684727295000111468(8)      $ & $5 \times 10^{-23}$ & $6\times 10^{-23}$ \\\hline
$1               $ & $    3.292867821434465922691(67)      $ & $4 \times 10^{-22}$ & $2\times 10^{-22}$ \\\hline
$2               $ & $    6.30388056744652609989(522)      $ & $2 \times 10^{-21}$ & $4\times 10^{-22}$ \\\hline
$3               $ & $    9.72732317270370501553(448)      $ & $5 \times 10^{-21}$ & $5\times 10^{-22}$ \\\hline
$4               $ & $    13.4812758360385893838(1489)     $ & $2 \times 10^{-20}$ & $2\times 10^{-21}$ \\\hline
$5               $ & $    17.5141323992530709259(6206)     $ & $3 \times 10^{-20}$ & $2\times 10^{-21}$ \\\hline
$6               $ & $    21.7909563917965158973(8744)     $ & $6 \times 10^{-20}$ & $3\times 10^{-21}$ \\\hline
$7               $ & $    26.286125156056810490(92289)     $ & $2 \times 10^{-19}$ & $7\times 10^{-21}$ \\\hline
$8               $ & $    30.979882837938369575(08213)     $ & $2 \times 10^{-19}$ & $8\times 10^{-21}$ \\\hline
$9               $ & $    35.856438766665971146(24181)     $ & $3 \times 10^{-19}$ & $9\times 10^{-21}$ \\\hline
\end{tabular}
\end{center}
\caption{Low-lying energy levels of the anharmonic quartic potential, obtained by diagonalization using level $p=13$ effective action. The parameters are  $M=\omega=1$, $g=48$, $L=5$, $\Delta=0.05$, $t=0.01$. For higher energy eigenvalues, absolute and relative errors $\Delta E_k$ and $\delta E_k$ are estimated by comparison with the diagonalization results obtained from higher-order effective actions, finer discretizations, larger space cutoffs, and lower values of the propagation time $t$.}
\label{tab:quarticspectrum}
\end{table}

Table~\ref{tab:quarticspectrum} gives low-lying energy eigenvalues of the anharmonic oscillator for a particular choice of the parameters of the potential and discretization parameters. In principle, one can achieve arbitrary precision by the use of appropriately chosen discretization parameters. Of course, for arbitrary precision calculations one has to use one of the software packages able to support such calculations. For example, we have used Mathematica \cite{mathematica} in order to be able to achieve high-precision results presented on the above graphs. The important conclusion is that even for very moderate values of discretization parameters, the use of higher-order effective actions leads to very small errors, which may be practically implemented with minimal computing resources.

The analysis of errors such as the one presented in Fig.~\ref{fig:quarticerr} is sufficient to estimate optimal values of discretization parameters. In general, for a desired numerical precision of energy eigenvalues, the optimal values of parameters are chosen so that all types of errors are approximately the same. The overall error is always dominated by the largest of all errors, and therefore it is optimal to have all errors of the same order of magnitude.

For specific calculations one can have additional constraints. For example, if one is interested only in energy eigenvalues, then the optimal parameters are obtained by minimizing all errors and minimizing the ratio $N=L/\Delta$, which corresponds to the size of the transition operator matrix $S=2N$ that needs to be numerically diagonalized. The minimization of $N$ is performed in order to minimize computation time needed for the diagonalization, which roughly scales as $N^3$.

On the other hand, if one is interested in details of energy eigenfunctions, then it might be necessary to have a fixed small value for the discretization step $\Delta$, which will allow all features of eigenstates to be visible. This is especially important for studies of higher energy eigenfunctions which e.g. have many nodes, and in order to study them it is necessary to have sufficient spatial resolution. In such case, the value of $\Delta$ is fixed and other parameters are chosen so as to minimize the errors to a desired value. For example, with the discretization step of the order $\Delta=10^{-3}$ we have been able to accurately calculate several hundreds energy eigenfunction of the quartic anharmonic oscillator.

Table~\ref{tab:doublewellspectrum} gives eigenvalues of the double-well potential, obtained from the quartic anharmonic potential (\ref{eq:quarticpot}) by setting the mass $M$ to some negative value. As can be seen, numerically obtained energy eigenvalues have the precision similar to the previous case of the quartic potential without symmetry breaking. The double well behavior of the potential does not present any obstacle in its numerical treatment by this method.

\begin{table}[!t]
\begin{center}
\begin{tabular}{|c|c|c|c|}
\hline
$k$ & $E_k$&$|\Delta E_k|$ & $\delta E_k$\\
\hline\hline
$\quad 0  \quad  $ & $  0.328826502590357561530(2)         $ & $7 \times 10^{-22}$ &$2 \times 10^{-21}$ \\\hline
$1               $ & $  1.41726810105965210733(23)         $ & $5 \times 10^{-21}$ &$4 \times 10^{-21}$ \\\hline
$2               $ & $  3.0819506284815341204(849)         $ & $3 \times 10^{-20}$ &$1 \times 10^{-20}$ \\\hline
$3               $ & $  5.019323060355788021(7990)         $ & $2 \times 10^{-19}$ &$4 \times 10^{-20}$ \\\hline
$4               $ & $  7.186203252338934478(3958)         $ & $5 \times 10^{-19}$ &$8 \times 10^{-20}$ \\\hline
$5               $ & $  9.54285734251209386(72421)         $ & $2 \times 10^{-18}$ &$2 \times 10^{-19}$ \\\hline
$6               $ & $ 12.06403774639116375(04211)         $ & $4 \times 10^{-18}$ &$4 \times 10^{-19}$ \\\hline
$7               $ & $ 14.7314279571006902(462590)         $ & $1 \times 10^{-17}$ &$7 \times 10^{-19}$ \\\hline
$8               $ & $ 17.5310745155383834(413592)         $ & $3 \times 10^{-17}$ &$2 \times 10^{-18}$ \\\hline
$9               $ & $ 20.4519281359123716(968554)         $ & $5 \times 10^{-17}$ &$3 \times 10^{-18}$ \\\hline
\end{tabular}
\end{center}
\caption{Low-lying energy levels of the double-well potential, obtained by diagonalization using level $p=18$ effective action. The parameters used:  $M=-1$, $\omega=1$, $g=12$, $L=16$, $\Delta=0.1$, $t=0.05$. The absolute and relative errors $\Delta E_k$ and $\delta E_k$ are estimated by comparison with the diagonalization results obtained from higher-order effective actions, finer discretizations, larger space cutoffs, and lower values of the propagation time $t$.}
\label{tab:doublewellspectrum}
\end{table}

Another situation in which one might be interested to keep the ratio $N=L/\Delta$, i.e. the size of the space-discretized evolution operator matrix as large as possible is when a large number of energy eigenlevels is needed. The number of energy eigenvalues that can be calculated by the diagonalization is limited by the size of the matrix $S=2N$. Usually the highest energy levels cannot be used due to the accumulation of numerical errors, and therefore one needs to have a matrix of sufficient size in order to study energy spectra. In such cases it is necessary to use highly optimized libraries for numeric diagonalization. We have implemented the effective actions as a C programming language code \cite{scl-speedup} and used LAPACK \cite{lapack} library for numeric diagonalization to calculate large number of energy eigenstates and eigenfunctions.

Even when one uses such a sophisticated tool, the highest eigenvalues cannot be used due to accumulation of numerical errors. In order to assess the quality of the obtained results for higher energy eigenstates, it is necessary to compare the numerical results with some known properties of the physical system. One such property is density of states, defined formally as
\begin{equation}
\rho(E)=\sum_{k=0}^\infty \delta(E-E_k)\, ,
\label{eq:dosdef}
\end{equation}
for a system with a discrete spectrum. This relevant physical quantity can be directly calculated using numerically obtained spectra. On the other hand, it can be also analytically calculated using semiclassical approximation. This approximation is valid at least in the high-energy region, and we can use it to assess the quality of our numerical results. In semiclassical approximation, the density of states is calculated as
\begin{equation}
\rho_{sc}(E)=\int\frac{\mathrm{d}^d\mathbf{x}\, \mathrm{d}^d\mathbf{p}}{(2\pi\hbar)^d}\,\delta(E-H(\mathbf{x},\mathbf{p}))\, .
\label{eq:dos-sc}
\end{equation}
replacing the discrete spectrum with a continuous distribution of energy defined by the classical Hamilton function $H(\mathbf{x},\mathbf{p})$. After integration over momenta, we obtain the well known result \cite{kleinertbook}
\begin{equation}
\rho_{sc}(E)=\frac{\left(\frac{M}{2\pi\hbar^2}\right)^{\frac{d}{2}}}{\Gamma\left(\frac{d}{2}\right)} \int\mathrm{d}^d\mathbf{x}\ \Theta(E-V(\mathbf{x}))\ (E-V(\mathbf{x}))^{\frac{d}{2}-1}\, ,
\label{eq:dos-sc-x}
\end{equation}
where $\Theta$ is the Heaviside step-function. For the quartic anharmonic potential (\ref{eq:quarticpot}) in $d=1$ the density of states can be expressed in terms of the complete elliptic integral of the first kind $K(k)=F(\pi/2, k)$ \cite{gradshteyn},
\begin{eqnarray}
\rho_{sc}(E)&=&\frac{\sqrt{2M/\pi^2\hbar^2}}{(M^2\omega^4/4+gE/6)^{1/4}}\ \times\nonumber\\
&& K\left(\sqrt{\frac{1}{2}-\frac{M\omega^2/4}{\sqrt{M^2\omega^4/4+gE/6}}}\, \right)\, .
\label{eq:dos-phi4}
\end{eqnarray}

\begin{figure}[!t]
\centering
\includegraphics[width=8.5cm]{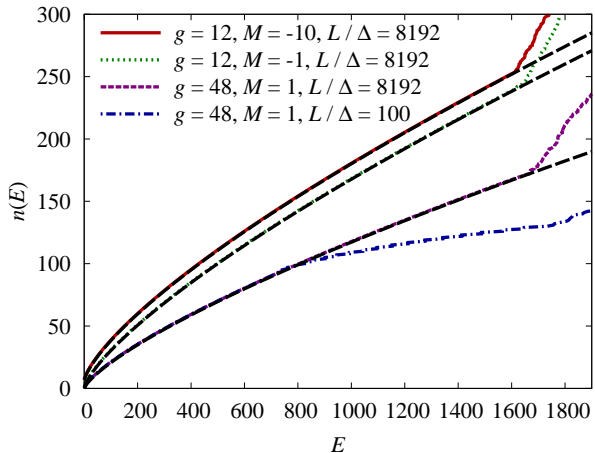}
\caption{(Color online) Cumulative distribution of the density of numerically obtained energy eigenstates for the quartic anharmonic and double-well potential, for $\omega=1$, $t=0.02$, $p=21$ and the following values of diagonalization parameters: $L=10$ for $g=12$ and $L=8$ for $g=48$. The discretization step is given on the graph by the value of $L/\Delta$, top to bottom. Long-dashed lines give corresponding semiclassical approximations for the cumulative density of states.}
\label{fig:cdos-phi4}
\end{figure}

In practical applications, especially in $d=1$, it might be difficult to compare directly semiclassical approximation for density of states and numerically obtained histogram for $\rho(E)$, since energy levels are usually not degenerated, so the spectrum is very sparse. In order to have sufficient statistics for a reasonable histogram, one has to use large value for bin size, and effectively the whole numerically available spectrum is reduced to just a few bins. For this reason, it is more instructive to study the cumulative density of states,
\begin{equation}
n(E)=\int_{V_{min}}^E\mathrm{d}E'\ \rho(E')\, ,
\label{eq:cdos-def}
\end{equation}
which counts the number of energy eigenstates smaller or equal to $E$. For quartic anharmonic oscillator the cumulative density of states is given by the above integral of the complete elliptic integral of the first kind, and can be calculated numerically. Fig.~\ref{fig:cdos-phi4} gives comparison of cumulative density of states calculated from our numerical diagonalization results and semiclassical approximation $n_{sc}(E)$.  As expected, the agreement is excellent up to very high values of energies, where numerical diagonalization eventually fails due to the finite number of calculated energy eigenvalues and effects of discretization. Such behavior can be improved by using finer discretization (smaller spacing size), as illustrated by two different discretization steps for $g=48$, $M=1$ in Fig.~\ref{fig:cdos-phi4}. Such analysis can be used to assess the obtained spectrum and determine the number of reliable energy eigenvalues. Typically we can achieve up to $10^4$ reliable energy eigenlevels with simulations on a single CPU. Note that the computer double precision accuracy of $10^{-16}$ imposes the limit on the maximal accessible energy eigenvalue $e^{-t E_{max}} \sim 10^{-16}$, i.e. $E_{max}\sim (16 \log 10)/t$. For Fig.~\ref{fig:cdos-phi4} we get $E_{max}\sim 1840$, which is above the limit imposed by the discretization used, as we can see from the graph.

\begin{figure}[!t]
\centering
\includegraphics[width=8.5cm]{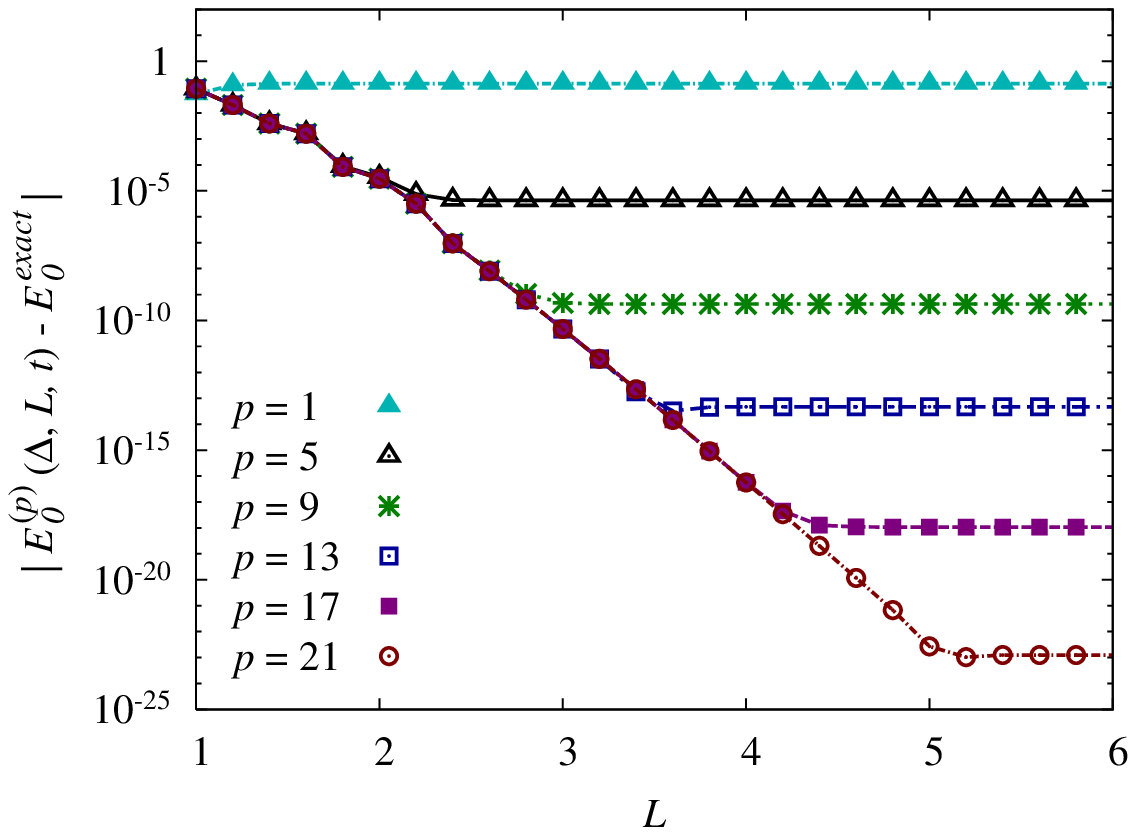}
\includegraphics[width=8.5cm]{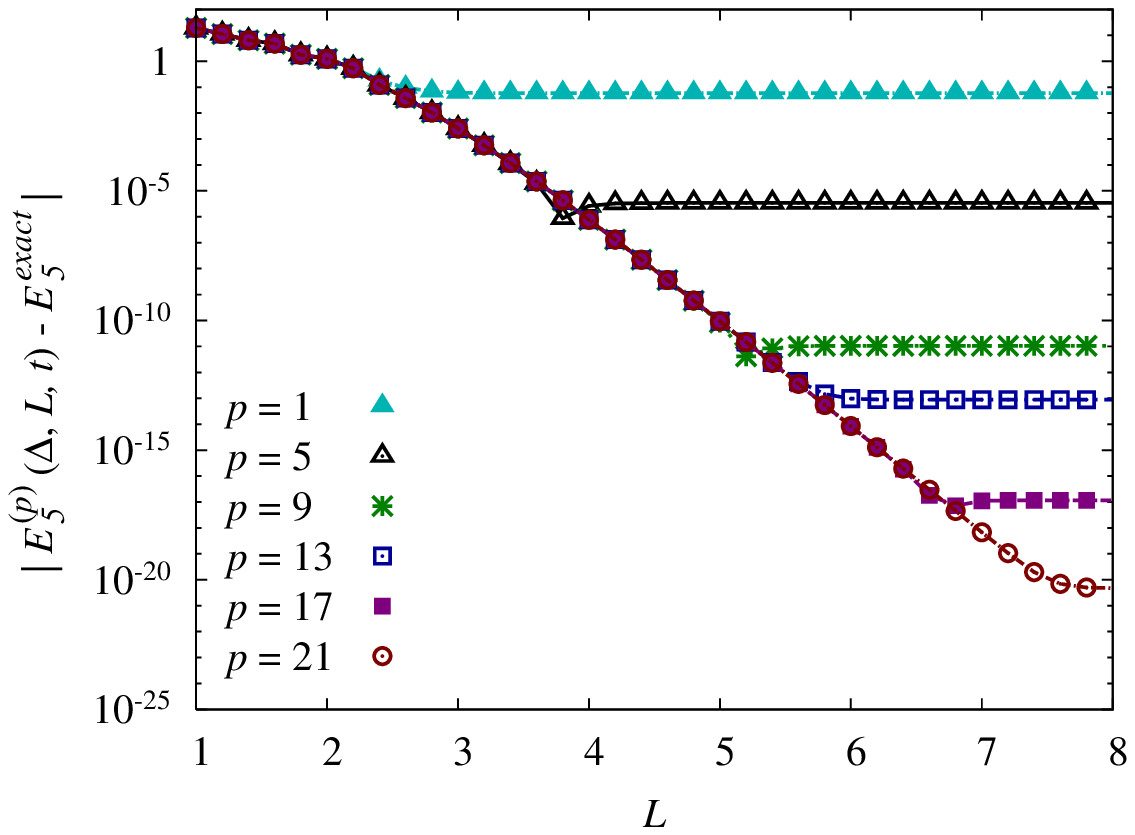}
\includegraphics[width=8.5cm]{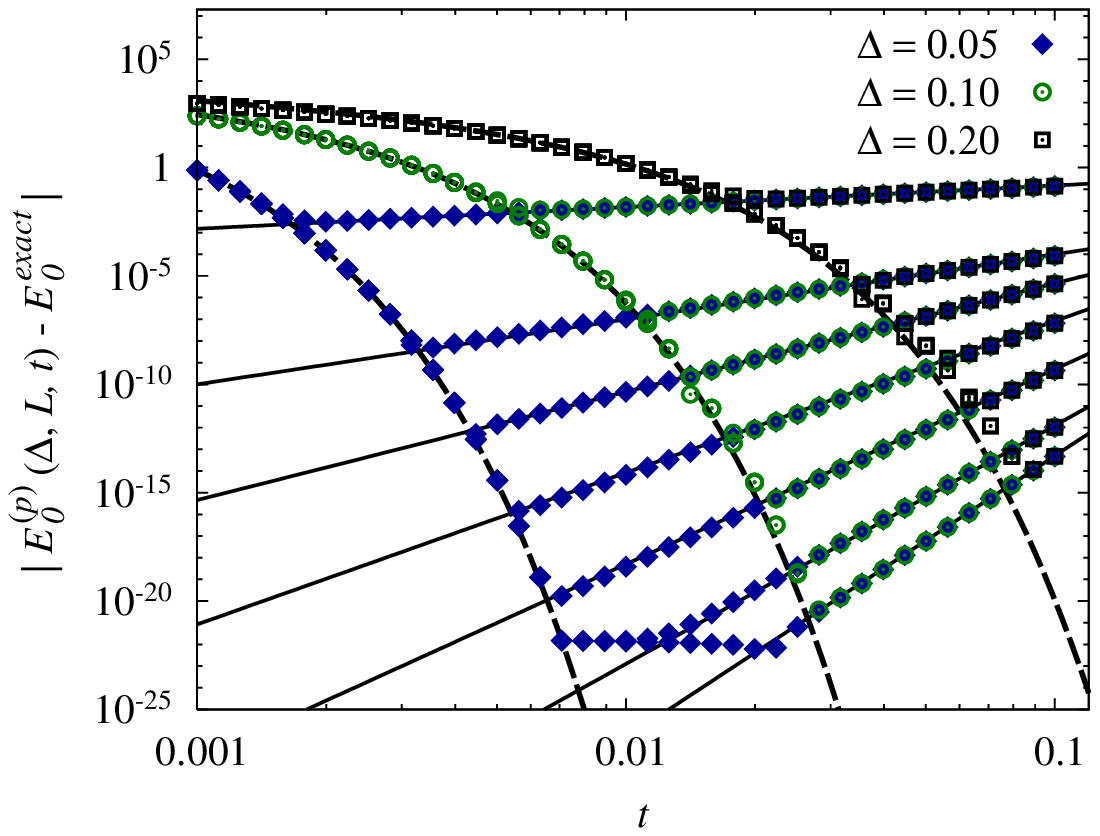}
\caption{((Color online) Deviations $|E_k^{(p)}(\Delta,L,t) -E_k^{exact}|$ as a function of $L$ for (a) $k=0$ and (b) $k=5$, and (c) as a function of $t$ for $k=0$, for the modified P\" oschl-Teller potential. Energy eigenvalues are obtained using effective action levels $p=1,5,9,13,17,21$ and $t=0.1$ in graphs (a) and (b), and $p=1,3,5,7,9,11,13$ and $L=5$ in graph (c), with the parameters $\alpha=0.5$, $\lambda=15.5$, $\Delta=0.02$. Dashed lines in the graph (c) correspond to the known discretization error \cite{pqseeo1}.}
\label{fig:mptdiff}
\end{figure}

In order to further demonstrate the applicability of the method, we also present numerical results for the modified P\" oschl-Teller model
\begin{equation}
V(x)=-\frac{\alpha^2}{2}\, \frac{\lambda (\lambda-1)}{\cosh^2\alpha x}\, ,
\label{eq:mptpot}
\end{equation}
which has only a finite set of discrete energy eigenlevels $E_k=-\alpha^2 (\lambda-1-k)^2/2$ for integer $k$ from the interval $0\leq k\leq \lambda-1$. Energy eigenvalues and eigenfunctions of this model are analytically known, and we will use them to further test our method. Effective actions to very high order are available also for this potential \cite{scl-speedup}, and we use them for numerical diagonalization of the evolution operator. Naturally, the diagonalization will give as many eigenvalues and eigenvectors as the size of the matrix $S$, but only the first few can be interpreted as bound states of the potential, according to the above condition  $0\leq k\leq \lambda-1$.

Fig.~\ref{fig:mptdiff}a gives the analysis of errors in the ground energy due to the space cutoff, while Fig.~\ref{fig:mptdiff}b gives the corresponding analysis of $L$-errors for numerical calculation of the energy level $E_5$. As we can see, the behavior of errors is the same as for the case of anharmonic oscillator, and we are again able to obtain high accuracy results. Fig.~\ref{fig:mptdiff}c gives the time dependence of errors in ground energy obtained by numerical diagonalization using different levels $p$ of effective actions. The scaling of errors proportional to $t^p$ is evident from the graph, as well as the discretization errors due to the finite discretization step $\Delta$. To ensure that the effective potential is bounded from below, in this case we have to remove higher-order powers of discretized velocity $\delta$ from the effective potential near $x=0$, since such terms have non-vanishing negative coefficients in the vicinity of $x=0$, due to a peculiar nature of the potential. In practical applications, one can use e.g. $p=1$ effective action (which does not depend on $\delta$) near $x=0$. As can be seen, this does not affect the obtained numerical results.

Table~\ref{tab:mptspectrum}a gives the obtained energy spectra for the modified P\"oschl-Teller potential with the parameters $\alpha=0.5$, $\lambda=15.5$. If necessary, the precision of obtained energy levels can be further increased by appropriately changing the discretization parameters. Contrary to the situation for anharmonic oscillator, where relative error of numerically calculated low-lying energy levels did not change significantly, here we see that the increase in the error is substantial. This is caused by the fact that this potential has only a small finite set of discrete bound states, so energy levels $k\sim 10$ correspond to the very top of the discrete spectrum. In practical applications such pathological situations are not encountered, but as we can see, even this can be dealt with by the proper choice of discretization parameters. The quality of numerically calculated eigenfunctions is assessed in Table~\ref{tab:mptspectrum}b, where we give a symmetric matrix of scalar products  $\langle\psi_k|\psi_l^{exact}\rangle$ of numerically calculated and analytic eigenfunctions. As we can see, the overlap between analytic and numeric eigenfunctions is excellent, and they are orthogonal with high precision, which is preserved even for higher energy levels. We have also verified that for parameters given in the caption of Table~\ref{tab:mptspectrum} and with the discretization step of the order $\Delta=10^{-3}$ eigenfunctions of all bound states can be accurately reproduced.

\onecolumngrid
\begin{center}
\begin{table}[!h]
\begin{center}
\begin{tabular}{|c|c|c|c|c|}
\hline
$k$ & $E_k$&$E_k^{exact}$&$|E_k-E_k^{exact}|$ &$\delta E_k$\\
\hline\hline
$\quad 0 \quad$  &     $-26.28125000000000000000000(174)$  &    $-26.28125$    &   $2\times 10^{-24}$ & $7\times 10^{-26}$\\\hline
$1$  &     $-22.781250000000000000000(28812)$  &    $-22.78125$    &   $3 \times 10^{-22}$ &$2 \times 10^{-23}$ \\\hline
$2$  &     $-19.53124999999999999999(736443)$  &    $-19.53125$    &   $3 \times 10^{-21}$ &$2 \times 10^{-22}$ \\\hline
$3$  &     $-16.5312499999999999999(6571136)$  &    $-16.53125$    &   $4 \times 10^{-20}$ &$2 \times 10^{-21}$ \\\hline
$4$  &     $-13.7812499999999999(8195897101)$   &    $-13.78125$    &  $2 \times 10^{-17}$ &$2 \times 10^{-18}$ \\\hline
$5$  &     $-11.28124999999999(398393103608)$   &   $-11.28125$     & $6 \times 10^{-15}$ &$6 \times 10^{-16}$ \\\hline
$6$  &     $-9.03124999999(8602255352218206)$    &   $-9.03125$      &  $2 \times 10^{-12}$ &$2 \times 10^{-13}$ \\\hline
$7$  &     $-7.031249999(773547728177905754)$    &   $-7.03125$      &  $3 \times 10^{-10}$ &$4 \times 10^{-11}$ \\\hline
$8$  &     $-5.2812499(74811672590174261082)$    &   $-5.28125$    & $3 \times 10^{-8}$ &$5 \times 10^{-9}$ \\\hline
\end{tabular}

\vspace*{4mm}
\begin{tabular}{|c|c|c|c|c|c|c|}
\hline
 &0&1&2&3&4&5\\
\hline
$\qquad 0\qquad$ &  $1-4.1\cdot 10^{-12}$& $1.2\cdot 10^{-13}$ & $2.8\cdot 10^{-6}$ & $7.4\cdot 10^{-14}$ & $3.0\cdot 10^{-7}$ & $7.9\cdot 10^{-14}$\\ \hline
$\qquad 1\qquad$ &  $1.2\cdot 10^{-13}$&  $1-1.1\cdot 10^{-11}$ & $2.1\cdot 10^{-13}$ & $4.5\cdot 10^{-6}$ & $4.6\cdot 10^{-14}$ & $6.5\cdot 10^{-7}$\\ \hline
$\qquad 2\qquad$ &  $2.8\cdot 10^{-6}$& $2.1\cdot 10^{-13}$ &  $1-2.2\cdot 10^{-11}$ & $1.3\cdot 10^{-13}$ & $5.9\cdot 10^{-6}$ & $2.2\cdot 10^{-13}$\\ \hline
$\qquad 3\qquad$ &  $7.4\cdot 10^{-14}$& $4.5\cdot 10^{-6}$ & $1.3\cdot 10^{-13}$ &  $1-3.5\cdot 10^{-11}$ & $3.1\cdot 10^{-13}$ & $6.9\cdot 10^{-6}$\\ \hline
$\qquad 4\qquad$ &  $3.0\cdot 10^{-7}$& $4.6\cdot 10^{-14}$ & $5.9\cdot 10^{-6}$ & $3.1\cdot 10^{-13}$ &  $1-4.8\cdot 10^{-11}$ & $7.9\cdot 10^{-13}$\\ \hline
$\qquad 5\qquad$ &  $7.9\cdot 10^{-14}$& $6.5\cdot 10^{-7}$ & $2.2\cdot 10^{-13}$ & $6.9\cdot 10^{-6}$ &  $7.9\cdot 10^{-13}$&  $1-5.8\cdot 10^{-11}$\\ \hline
\end{tabular}
\end{center}
\caption{(a) Low-lying energy levels of the modified P\" oschl-Teller potential, obtained by diagonalization using level $p=21$ effective action. The parameters used:  $\alpha=0.5$, $\lambda=15.5$, $L=5$, $\Delta=0.02$, $t=0.1$. (b) Symmetric table of scalar products  $\langle\psi_k|\psi_l^{exact}\rangle$ of numerically calculated and analytic eigenstates for $k,l=0,1,2,3,4,5$.}
\label{tab:mptspectrum}
\end{table}
\end{center}
\twocolumngrid

\section{Numerical results for $d=2$ models}
\label{sec:d2}

In this section we illustrate the application of the numerical method based on the diagonalization of transition amplitudes on two $d=2$ models. The first model is the anharmonic oscillator
\begin{eqnarray}
&&\hspace*{-5mm}V(x,y)=\frac{1}{2}M\, (\omega_\perp^2-\Omega^2)\,  (x^2+y^2)+\frac{g}{24}(x^2+y^2)^2\nonumber\\
&&=\frac{1}{2}M\omega_\perp^2 (1-r^2)\, (x^2+y^2)+\frac{g}{24}(x^2+y^2)^2\, ,
\label{eq:2dquarticpot}
\end{eqnarray}
which is used for a description of the trapping potential used in a recent experiment with fast-rotating Bose-Einstein condensate of $^{87}$Rb atoms \cite{dalibard2004,dalibard2005, pelster2007}. The harmonic frequency of the trapping potential is partially compensated by the rotation frequency $\Omega$. The small quartic anharmonicity is used in order to allow the condensate to be rotated at the critical frequency $\Omega=\omega_\perp$, and still to remain confined. The ratio  $r=\Omega/\omega_\perp$ is used to express rotation frequency in suitable units of harmonic frequency $\omega_\perp$.

\begin{figure*}[!t]
\centering
\includegraphics[width=8cm]{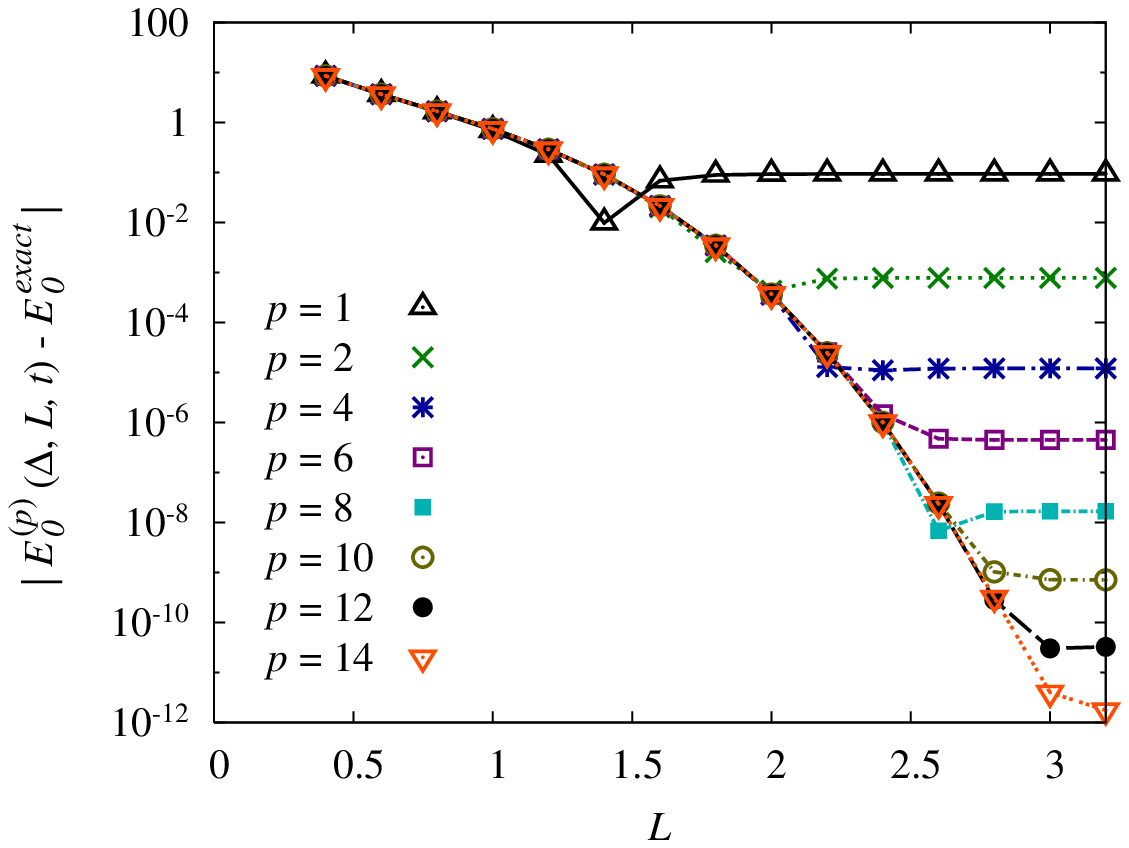}
\includegraphics[width=8cm]{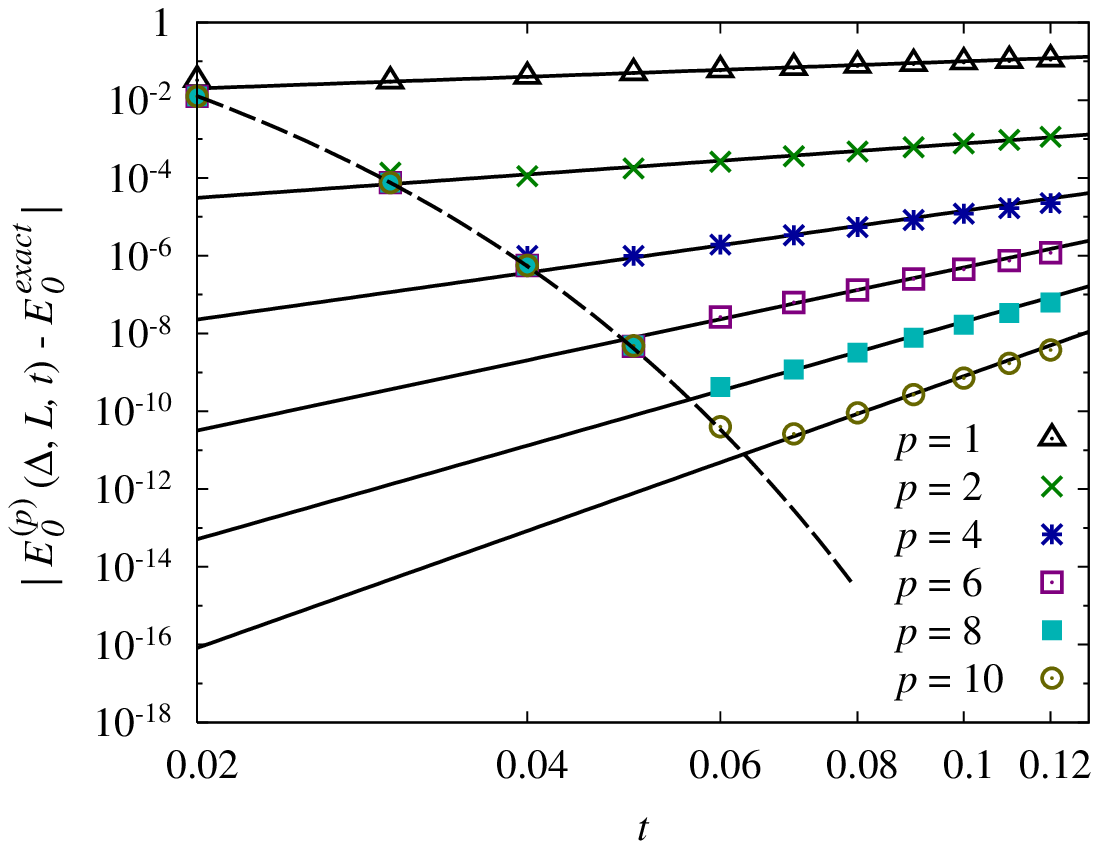}
\caption{(Color online) (a) Deviations from the ground energy $|E_0^{(p)}(\Delta,L,t) -E_0|$ as a function of the space cutoff $L$ and (b) as a function of the time $t$ for a critically rotating gas of $^{87}$Rb atoms in a $d=2$ anharmonic trap with $g\approx  2\cdot 10^3 g_{exp}$. The discretization parameters are $\Delta=0.2$, $t=0.1$ on the graph (a), and $L=3.2$ on the graph (b). Deviations are calculated using the ground energy $E_0=1.47714975357799(4)$ obtained with $p=21$ effective action. The dashed line in graph (b) corresponds to the known discretization error \cite{pqseeo1}.}
\label{fig:BECerr}
\end{figure*}
\begin{table*}[!t]
\begin{center}
\begin{tabular}{|c|c|c|c|}
\hline
$k$ &$E_k/\hbar\omega_\perp$, $r=0$, $g=g_{exp}$& $E_k/\hbar\omega_\perp$ $r=1.05$, $g=g_{exp}$ & $E_k/\hbar\omega_\perp$, $r=1.05$, $g=10^3\, g_{exp}$ \\
\hline\hline
$\quad 0  \quad  $ & $  1.0009731351803  $ & -1.1279858856602 &  1.1287297831435 \\\hline
$1               $ & $    2.0029165834022      $ & -1.1169327267787 & 2.6161348497834  \\\hline
$2               $ & $   2.0029165834022      $ & -1.1169327267787  & 2.6161348497834 \\\hline
$3               $ & $    3.0058275442161      $ & -1.0842518375067  & 4.3476515279810 \\\hline
$4               $ & $    3.0058275442161     $ & -1.0842518374840  & 4.3476515279812 \\\hline
$5               $ & $    3.0067964582067     $ & -1.0311383813261  & 4.6528451852013 \\\hline
$6               $ & $    4.0097032385903     $ & -1.0311383813261  & 6.2704552903671 \\\hline
$7               $ & $    4.0097032385903    $ & -0.95910186300510  & 6.2704552903671 \\\hline
$8               $ & $    4.0116368851078     $ & -0.95910186300478  & 6.7589882491411 \\\hline
$9               $ & $    4.0116368851078     $ & -0.86968170695135  & 6.7589882491412 \\\hline
\end{tabular}
\end{center}
\caption{Low-lying energy levels of a rotating gas of $^{87}$Rb atoms in a $d=2$ anharmonic trap, obtained using the level $p=21$ effective action. The discretization parameters are $L=14$, $\Delta=0.14$, and $t=0.2$.}
\label{tab:BECspectrum}
\end{table*}

The typical values of parameters used in the experiment are $\omega_\perp=2\pi\cdot 64.8$ Hz and $g=g_{exp}=1.56\cdot 10^{-10}$ J/m$^4$. In Fig.~\ref{fig:BECerr} we have used much larger quartic coupling $g\approx  2\cdot 10^3 g_{exp}$ in order to increase the non-harmonic effects of the potential. Also, the graphs in Fig.~\ref{fig:BECerr} are calculated for critical rotation ($r=1$), where the potential is reduced to a pure quartic interaction. The analysis of errors is very similar as in the one-dimensional cases we studied in the previous section. The dependence of ground energy errors on the space cutoff $L$ is shown in Fig.~\ref{fig:BECerr}a, and we see the usual saturation of errors for sufficiently large values of $L$. The saturated value rapidly decreases (by several orders of magnitude) as we increase the level $p$ of the effective action used to calculate space-discretized matrix of the evolution operator. Fig.~\ref{fig:BECerr}b shows the time dependence of ground energy errors, which are found to fully agree with the scaling law $t^p$ for sufficiently fine discretization. Again, the discretization errors basically conform to the universal dependence given in Eq.~(17) of our previous paper \cite{pqseeo1}. The dimensionality of the system introduces an overall multiplicative factor of 2, and the additional factor of 2 in the $\cosh$ term.

\begin{figure*}[!t]
\centering
\includegraphics[height=5.5cm]{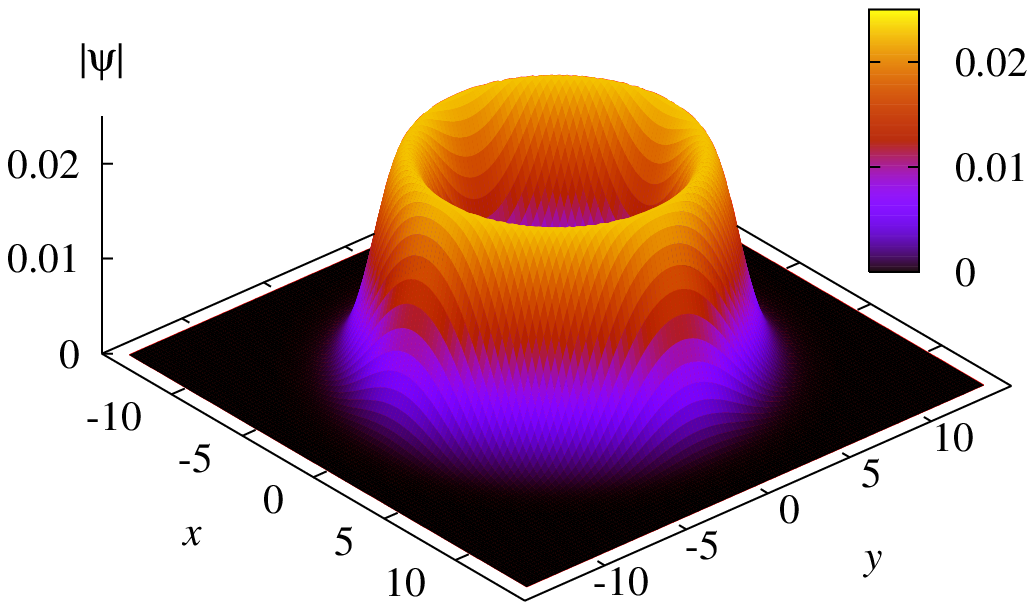}
\includegraphics[height=5.5cm]{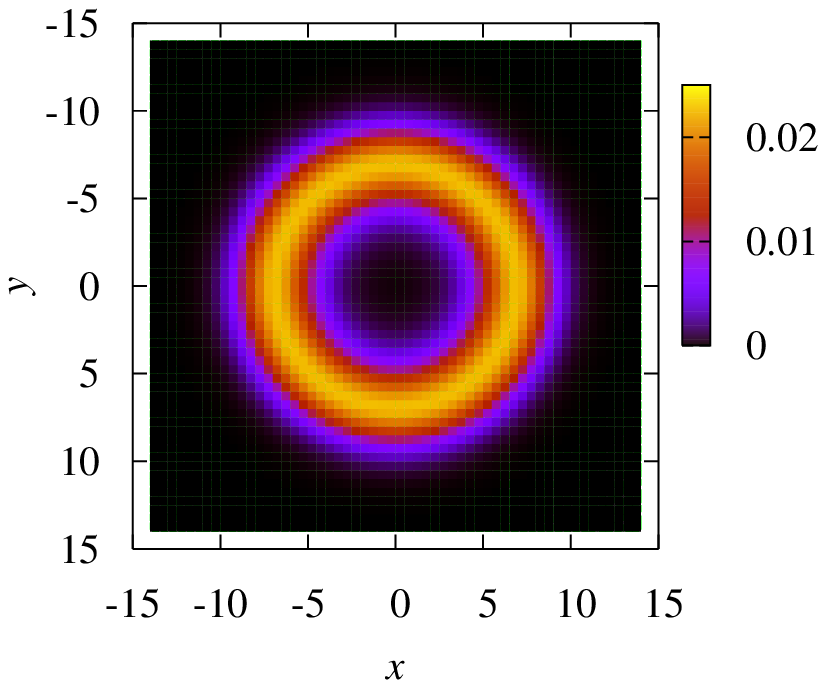}
\caption{(Color online) Ground state (as 3-D plot on the left, and as a density plot on the right) of a rotating gas of $^{87}$Rb atoms in a $d=2$ anharmonic trap obtained using  $p=21$ effective action. The parameters are $r=1.05$, $g=g_{exp}$, $L=20$, $\Delta=0.25$, $t=0.2$.}
\label{fig:BECE0}
\end{figure*}
\begin{figure}[!b]
\centering
\includegraphics[width=4.2cm]{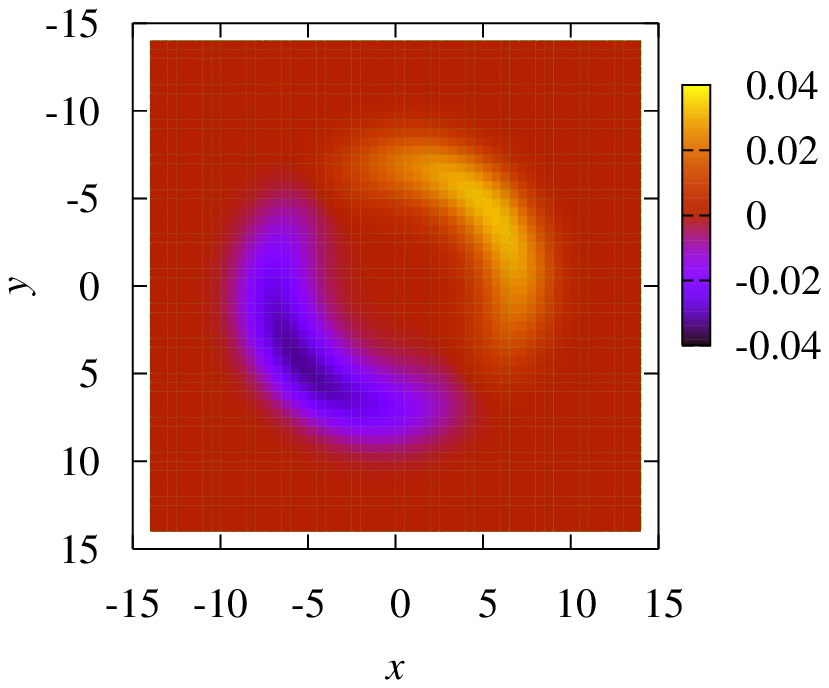}
\includegraphics[width=4.2cm]{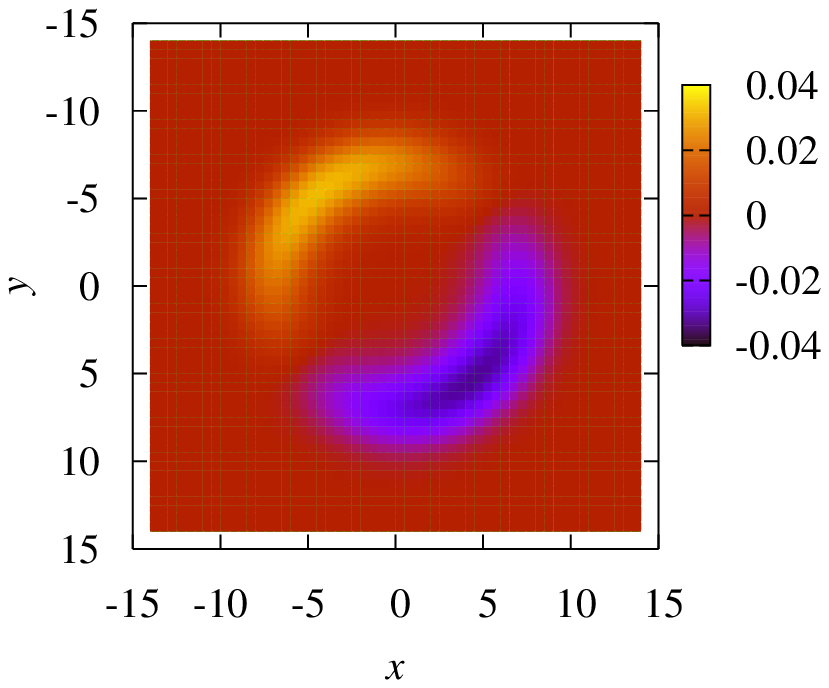}
\includegraphics[width=4.2cm]{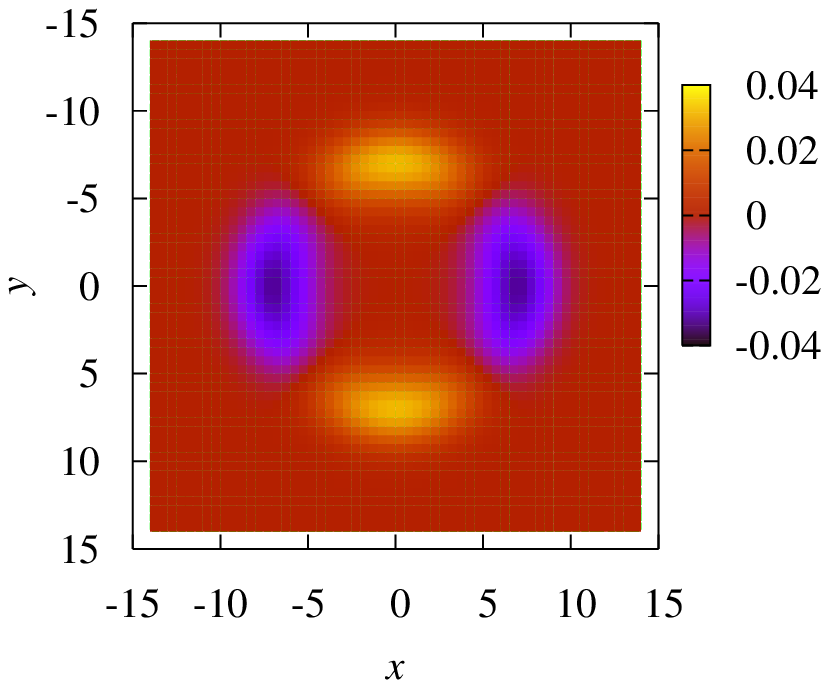}
\includegraphics[width=4.2cm]{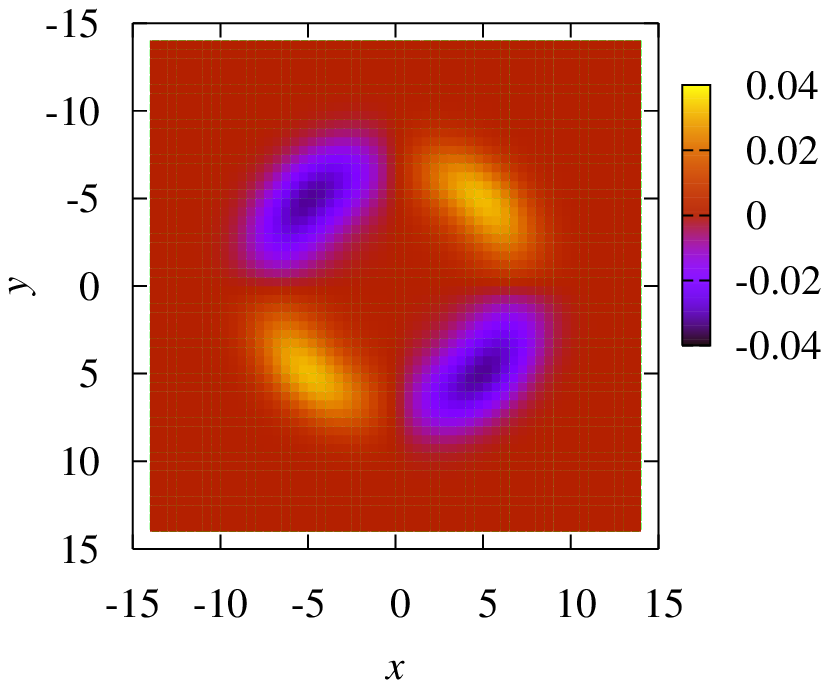}
\caption{(Color online) Density plots of level $k=1,2,3,4$ eigenstates of a rotating gas of $^{87}$Rb atoms in a $d=2$ anharmonic trap obtained using  $p=21$ effective action. The parameters are $r=1.05$, $g=g_{exp}$, $L=20$, $\Delta=0.25$, $t=0.2$.}
\label{fig:BECphi1234}
\end{figure}

Table~\ref{tab:BECspectrum} gives the numerically obtained energy eigenvalues for different sets of parameters of the potential: non-rotating system, system with overcritical rotation ($r=1.05$), and system with overcritical rotation, but with significantly larger anharmonicity ($g=10^3\, g_{exp}$). From the analysis of discretization errors and errors related to the use of a chosen effective action level $p$, we can estimate the errors in found energy eigenvalues to be of the order $10^{-15}$, where we express energy in units of $\hbar\omega_\perp$. The results in the Table~\ref{tab:BECspectrum} are obtained by numerical diagonalization based on the C SPEEDUP code \cite{scl-speedup} and the use of the LAPACK \cite{lapack} library. The estimated error in energy eigenvalues is smaller than the (relative) error which can be achieved in typical C simulations, which is of the order $10^{-14}$. This is easily verified, since for several different values of discretization parameters we get the same stable results shown in the table. Therefore, this table gives certain digits in all energy eigenvalues, and the error can be cited as implicit (half of the last digit). This is good example for practical applications, where we have managed to eliminate all types of errors below the limit that can be seen due to inherent numerical errors of computer simulation. However, if such complete elimination of errors is not possible due to the limitations in computer memory or computation time, the analysis of errors presented in Fig.~\ref{fig:BECerr} allows us to reliably estimate numerical errors in energy eigenvalues.

Fig.~\ref{fig:BECE0} shows the numerically obtained ground state for this two-dimensional potential for the case of  overcritical rotation. The ground state has the expected Mexican hat shape. The figure gives a three-dimensional plot of the ground state on the left, and the corresponding density plot on the right, with values of the wave function mapped to colors. Fig.~\ref{fig:BECphi1234} gives density plots of $k=1,2,3,4$ eigenfunctions for the same values of parameters. The discretization is sufficiently fine ($\Delta=0.25$) in rescaled dimensionless units) so that all features of calculated eigenfunctions are clearly visible.

The numerical study of this example related to Bose-Einstein condensation is chosen as an example where ground state eigenfunction is necessary with high resolution in order to calculate e.g. time-of-flight absorption graphs \cite{progress} and to study formation and evolution of vortices in the condensate. In addition to this, large numbers of accurate eigenstates are needed for calculation of the condensation temperature, condensate fraction, and other static and dynamic properties of the condensate. For this reason, it is necessary to assess numerically obtained eigenstates and use only reliable ones in further calculation. As in the one-dimensional case, we will calculate the density of states $\rho_{sc}(E)$ in semiclassical approximation, and use it as a criterion for the reliability of high-energy eigenstates. In $d=2$, the density of states is given by a simple formula
\begin{equation}
\rho_{sc}(E)=\frac{M}{2\pi\hbar^2} \int\mathrm{d}\mathbf{x}\ \Theta(E-V(\mathbf{x}))\, .
\label{eq:dos-BEC-sc-x}
\end{equation}
For the quartic anharmonic potential (\ref{eq:2dquarticpot}) the density of states can be analytically calculated
\begin{eqnarray}
&&\hspace*{-15mm}\rho_{sc}(E)=\frac{M}{2\hbar^2}\ \left(-\frac{6M\omega_\perp^2\, (1-r^2)}{g}\ +\right.\nonumber\\
&&\left.\sqrt{\left(\frac{6M\omega_\perp^2\, (1-r^2)}{g}\right)^2+\frac{24E}{g}}\ \right)\, ,
\label{eq:dos--units-BEC}
\end{eqnarray}
or, in dimensionless units used in all numerical calculations,
\begin{equation}
\rho_{sc}(E)=-\frac{3\, (1-r^2)}{g}+\sqrt{\frac{9\, (1-r^2)^2}{g^2}+\frac{6E}{g}}\, .
\label{eq:dos-BEC}
\end{equation}

\begin{figure}[!t]
\centering
\includegraphics[width=7cm]{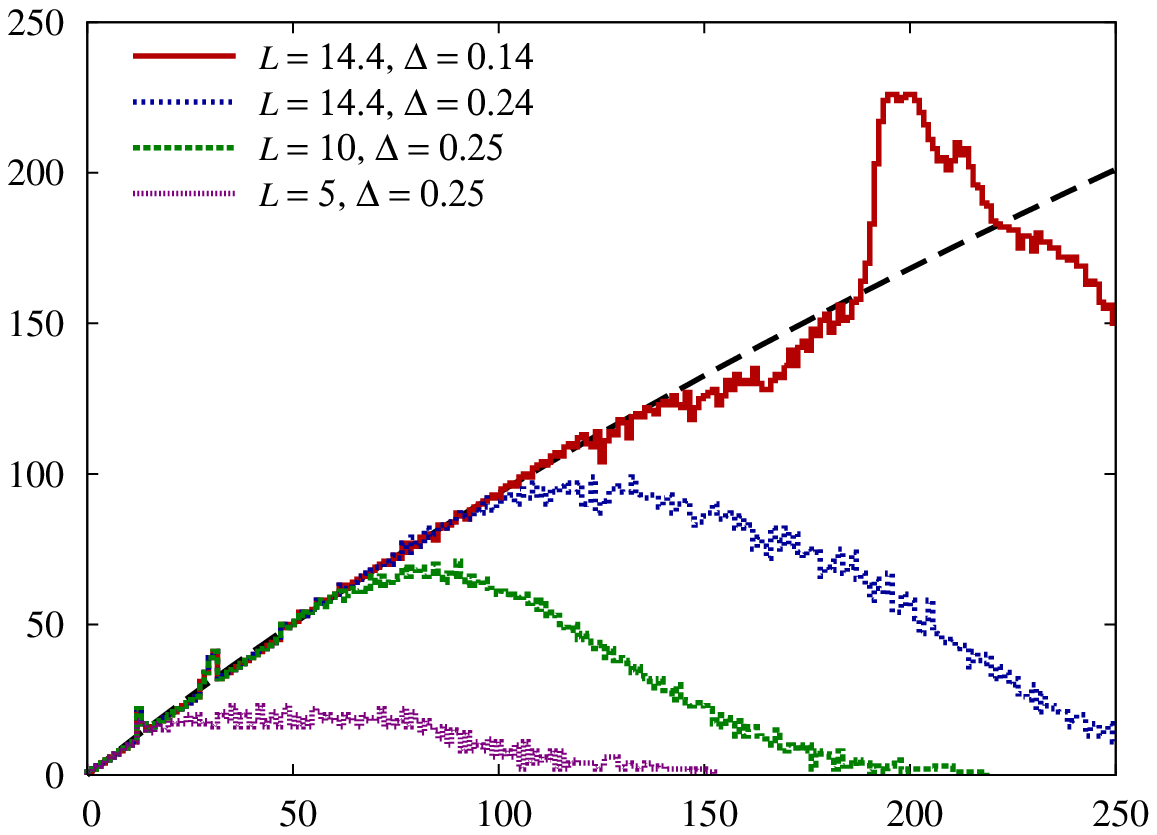}
\includegraphics[width=7cm]{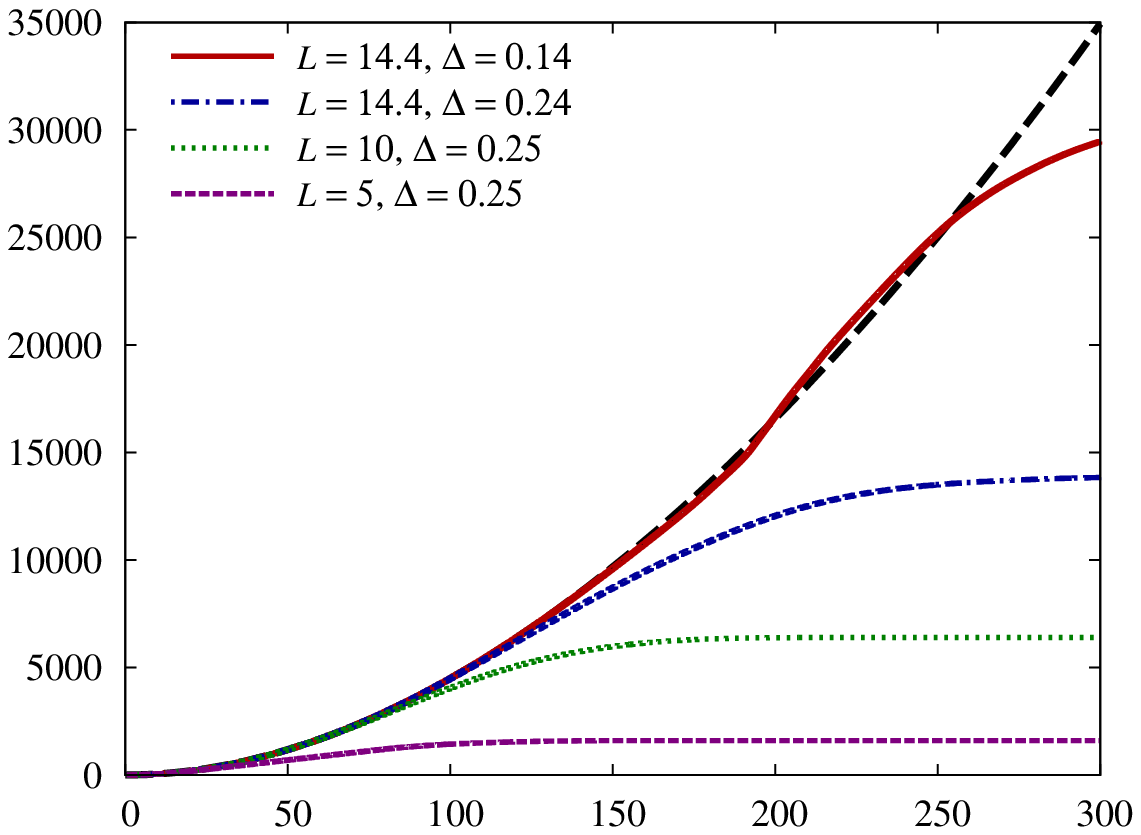}
\caption{(Color online) (a) Distribution of the density of numerically obtained energy eigenstates and (b) cumulative distribution of the density of numerically obtained energy eigenstates for non-rotating gas of $^{87}$Rb atoms in a $d=2$ anharmonic trap, calculated with the level $p=21$ effective action. The parameters are $r=0$, $g=g_{exp}$, $t=0.2$, while discretization parameters are given on the graph, corresponding to the curves top to bottom. Long-dashed lines on both graphs give the corresponding semiclassical approximations.}
\label{fig:dos-BEC}
\end{figure}

Fig.~\ref{fig:dos-BEC}a shows the comparison of semiclassical approximation for the density of states, and the histogram for numerically obtained energy eigenvalues of the potential (\ref{eq:2dquarticpot}). Due to the high degeneracy of energy eigenstates in $d=2$, the histogram of numerically found energy levels contains enough statistics over the whole region of energies, and therefore can be used for assessment of the quality of numerical spectra. As we see, the agreement is better and better when we use finer space discretization. Depending on the needed number of energy levels and maximal value of the energy considered to be relevant for the calculation we can choose appropriate values of discretization parameters that will provide reliable numerical results up to desired energy value. For example, for the choice of discretization parameters $L=14$, $\Delta=0.14$, we can reliably use energy levels up to $E\approx 120\, \hbar\omega_\perp$.

Fig.~\ref{fig:dos-BEC}b shows the comparison of cumulative density of states $n(E)$ calculated for numerically obtained results and in semiclassical approximation, by integrating the expression (\ref{eq:dos-BEC}), which can be calculated analytically. The comparison of numerical and semiclassical cumulative density of states in Fig.~\ref{fig:dos-BEC}b verifies our conclusions from Fig.~\ref{fig:dos-BEC}a, and again sets the same limit of reliable energy levels for chosen discretization parameters.

\begin{table}[!t]
\begin{center}
\begin{tabular}{|c|c|c|c|c|}
\hline
$i$ &$V_{i0}$& $a_i$ & $b_i$ & $c_i$\\
\hline\hline
$x$ &  100 & 1.56 & -0.61 & 0.32 \\ \hline
$y$ &  100 & 0.69 & -0.12 & 0.03 \\ \hline
$xy$ &  100 & -1.00 & 0.25 & 0.08 \\ \hline
\end{tabular}
\end{center}
\caption{Parameters of the sextic potential (\ref{eq:sextic}).}
\label{tab:sextic}
\end{table}

\begin{figure}[!b]
\centering
\includegraphics[width=7cm]{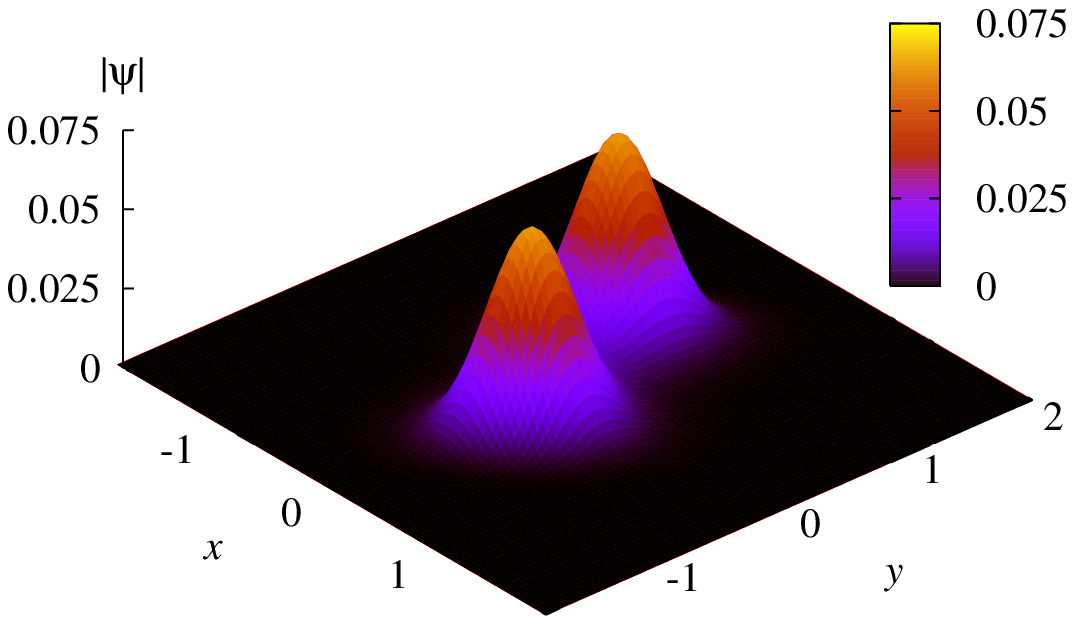}
\includegraphics[width=7cm]{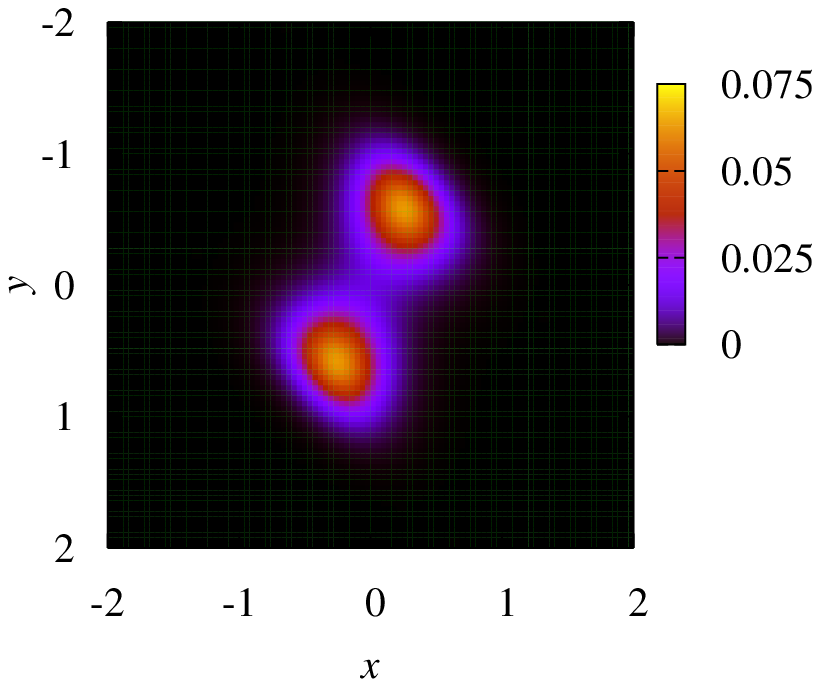}
\caption{(Color online) Ground state (as a 3-D and as a density plot) of a sextic anharmonic potential, obtained by diagonalization using the level $p=21$ effective action. The parameters of the potential are given in the text. The diagonalization parameters: $L=4$, $\Delta=0.04$, $t=0.01$.}\label{fig:phi6E0}
\end{figure}

\begin{figure}[!t]
\centering
\includegraphics[width=4.2cm]{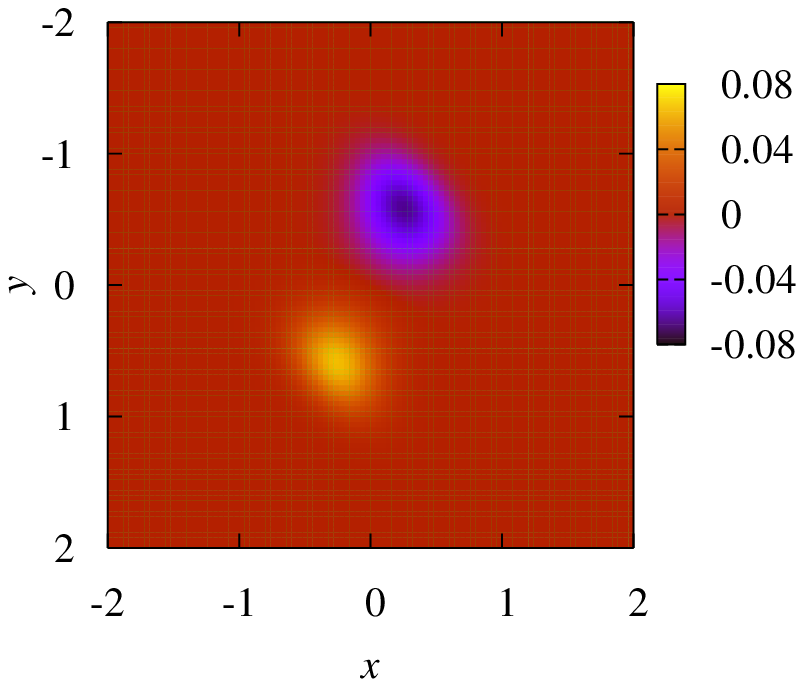}
\includegraphics[width=4.2cm]{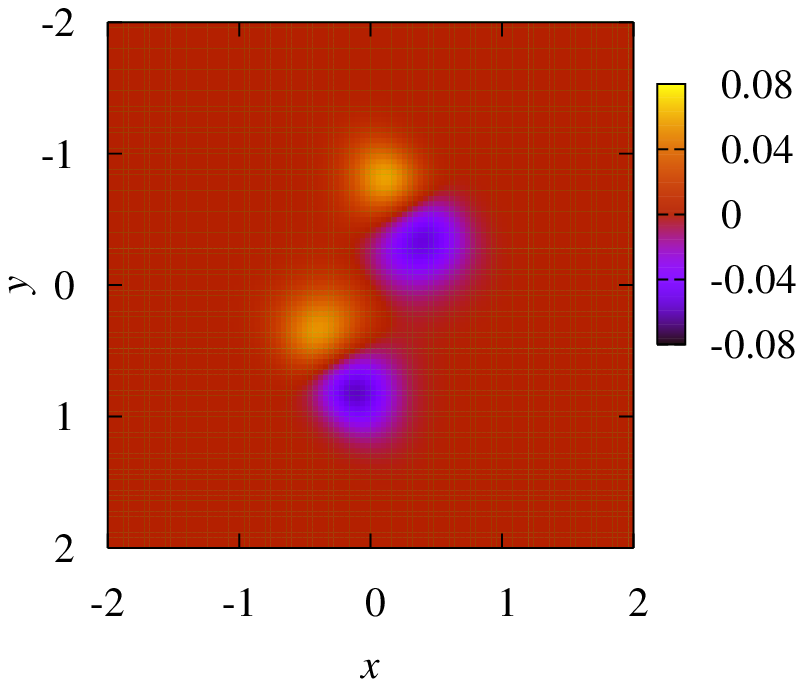}
\includegraphics[width=4.2cm]{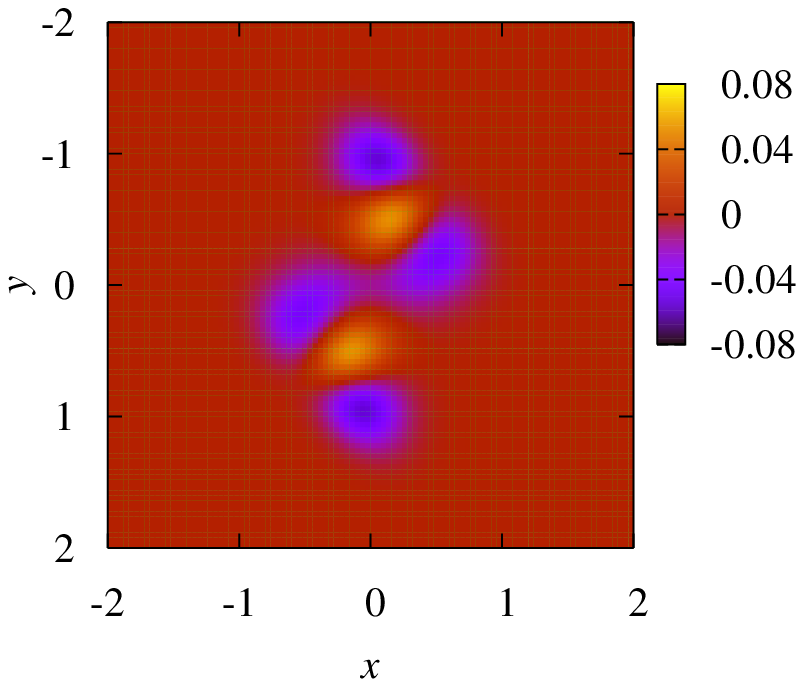}
\includegraphics[width=4.2cm]{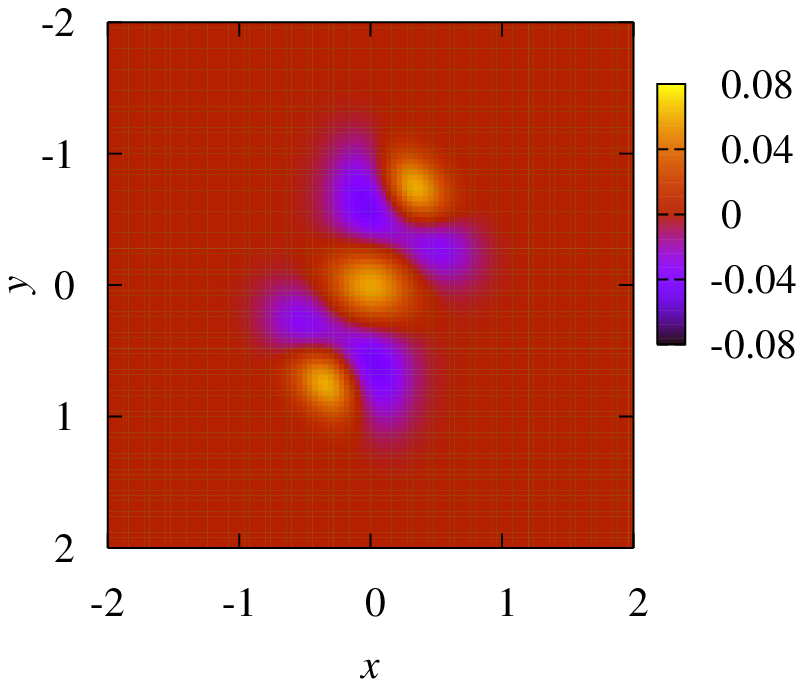}
\caption{(Color online) Density plots of level $k=1,3,7,8$ eigenstates of a sextic anharmonic potential, obtained by diagonalization using the level $p=21$ effective action. The parameters of the potential are given in the text. The diagonalization parameters: $L=4$, $\Delta=0.04$, $t=0.01$.}
\label{fig:phi6phi1378}
\end{figure}

The second two-dimensional model we have studied numerically is a sextic anharmonic oscillator,
\begin{equation}
V(x,y)=V_x(x)+V_y(y)+V_{xy}(x-y)\, ,
\label{eq:sextic}
\end{equation}
where $V_i(x)=V_{i0}\, (a_i x^2+b_i x^4+c_i x^6)$. The values of the coefficients used are given in Table~\ref{tab:sextic}. The study of this potential is motivated by Ref.~\cite{seligman}, where it has been used to investigate the transition from regular to chaotic classical motion. Fig.~\ref{fig:phi6E0} shows the numerically obtained ground state for this two-dimensional potential, as a three-dimensional plot on the left, and as a density plot on the right. Fig.~\ref{fig:phi6phi1378} gives density plots of $k=1,3,7,8$ eigenfunctions for the same values of parameters. The discretization is sufficiently fine ($\Delta=0.04$) so that we can resolve all details in the presented eigenstates.

We have demonstrated that the presented approach can be successfully used for numerical studies of lower-dimensional models. Note that in $d=3$ the complexity of the algorithm and sizes of matrices to be diagonalized may practically limit the applicability to the calculation of only low-lying energy levels. Also, in this case it might be difficult to numerically obtain three-dimensional eigenfunctions on finer grids, since even moderate grids with 50-100 points in one dimension would require exact diagonalization of extremely large matrices.

At the end, let us compare the complexity of the presented approach and direct diagonalization of the space-discretized Hamiltonian, as well as finite-element methods. The main difference in the complexity of algorithms is related to the exponential growth in the size of analytic expressions for the effective potential with the increase of the level $p$, as discussed in Ref.~\cite{b}. Therefore, the required CPU time for construction of the matrix to be diagonalized in the presented approach grows exponentially with the level $p$, while in other methods the construction of such a matrix does not require a significant amount of time. However, the time for exact diagonalization far outweighs the time needed for construction of even large matrices with moderate levels $p$ of the order 10-20. The significant benefit of practically eliminating errors associated with the time of propagation therefore fully justifies the use of the effective action approach. Of course, in practical applications one has to study the complexity of the algorithm and to choose the optimal level $p$ which will sufficiently reduce the errors, while keeping the complexity of the calculation on the acceptable level.

\section{Conclusions}
\label{sec:conclusions}

In this paper we have presented a substantial improvement of previously introduced method \cite{sethia} for study or properties of quantum systems using numeric diagonalization of the space-discretized evolution operator. This approach allows exact numeric calculation of a large number of energy eigenvalues and eigenstates of the system. Our previous paper \cite{pqseeo1} has presented detailed analysis of all types of discretization errors inherent to this method, which were not analyzed completely before.

This paper resolves a key problem in practical applications of this approach: accurate calculation of transition amplitudes, matrix elements of the space-discretized evolution operator. Using recently introduced effective action approach \cite{b} that gives systematic short-time expansion of the evolution operator, we can analytically calculate matrix elements of the evolution operator with high precision. This enables high precision calculation of energy eigenvalues and eigenstates, as was shown in this paper.

The derived analytical estimates for all types of errors, including errors due to the approximative calculation of transition amplitudes, provide us with a way to choose optimal discretization parameters and to reduce overall errors in energy eigenvalues and eigenstates for many orders of magnitude, as was demonstrated for several one- and two-dimensional models. We have shown that numerical diagonalization of the space-discretized evolution operator can be successfully applied for studies of many interesting lower dimensional models. Due to the superior behavior of discretization and other errors in this method compared to methods based on diagonalization of the discretized Hamilton operator and related methods, the presented approach is a method of choice for numerical studies of lower-dimensional physical systems. The authors are already using this approach for numerical investigation of properties of fast rotating Bose-Einstein condensates \cite{progress}, and plan to use it for the treatment of dilute quantum gases in a disordered environment. Another interesting line of research would be combining the present method with the density matrix renormalization group (DMRG) approach \cite{d, scholl}.

\section*{Acknowledgments}
This work was supported in part by the Ministry of Science and Technological Development of the Republic of Serbia, under project No. OI141035 and bilateral project PI-BEC funded jointly with the German Academic Exchange Service (DAAD), and the European Commission under EU Centre of Excellence grant CX-CMCS. Numerical simulations were run on the AEGIS e-Infrastructure, supported in part by FP7 projects EGEE-III and SEE-GRID-SCI.

\end{document}